\documentclass[english]{aastex}
\usepackage[T1]{fontenc}
\usepackage{amsbsy}
\usepackage{amstext}
\usepackage{amssymb}
\usepackage{graphicx}

\makeatletter
\slugcomment{}
\shorttitle{}
\shortauthors{}

\makeatother

\usepackage{babel}
\begin{document}

\title{The Multi-Species Farley-Buneman Instability in the Solar Chromosphere}

\author{Chad A. Madsen}

\email{cmadsen@bu.edu}

\author{Yakov S. Dimant}

\author{Meers M. Oppenheim}

\affil{Center for Space Physics, Boston University, 725 Commonwealth Ave.,
Boston, MA 02215 }

\and{}

\author{Juan M. Fontenla}

\affil{Laboratory for Atmospheric and Space Physics, University of Colorado
at Boulder, 1234 Innovation Dr., Boulder, CO 80303}
\begin{abstract}
Empirical models of the solar chromosphere show intense electron heating
immediately above its temperature minimum. Mechanisms such as resistive
dissipation and shock waves appear insufficient to account for the
persistence and uniformity of this heating as inferred from both UV
lines and continuum measurements. This paper further develops the
theory of the Farley-Buneman Instability (FBI) which could contribute
substantially to this heating. It expands upon the single ion theory
presented by Fontenla (2005) by developing a multiple ion species
approach that better models the diverse, metal-dominated ion plasma
of the solar chromosphere. This analysis generates a linear dispersion
relationship that predicts the critical electron drift velocity needed
to trigger the instability. Using careful estimates of collision frequencies
and a one-dimensional, semi-empirical model of the chromosphere, this
new theory predicts that the instability may be triggered by velocities
as low as 4 km s$^{\text{-1}}$, well below the neutral acoustic speed.
In the Earth's ionosphere, the FBI occurs frequently in situations
where the instability trigger speed significantly exceeds the neutral
acoustic speed. From this, we expect neutral flows rising from the
photosphere to have enough energy to easily create electric fields
and electron Hall drifts with sufficient amplitude to make the FBI
common in the chromosphere. If so, this process will provide a mechanism
to convert neutral flow and turbulence energy into electron thermal
energy in the quiet Sun.
\end{abstract}


\section{Introduction}

Just above the temperature minimum in the solar chromosphere, the
plasma temperature rises steeply by over 2000 K. This region of intense
heating explained the continuum and line emissions in the quiet-Sun
chromosphere, but the energy source remains unexplained (Athay 1966).
Detailed semi-empirical models of this region by Vernazza et al. (1981)
and Fontenla et al. (1991; 1993; 2009) also predict the enhanced temperature
region shown in Figure 1, but did not explain the mechanisms sustaining
it. 

Researchers have proposed a number of mechanisms to explain the heating
but none have proven compelling. Carlsson \& Stein (1992) suggested
that acoustic shocks were responsible; however, the predicted temporary
variations in the Ca II K line profile remain unobserved (Carlsson
2007). Campos \& Mendes (1995) proposed resistive heating due to steady
electric currents or MHD wave dissipation as possible sources of the
enhanced temperature region; however, chromospheric plasma does not
appear to have a sufficiently high conductivity to produce this region
via classical Joule heating (Socas-Navarro 2007).

Fontenla (2005) and Fontenla et al. (2008) suggested that plasma turbulence
due to the Farley-Buneman Instability (FBI) can heat some layers of
the chromosphere. They argue that convective motions from the photosphere
will drag ions across the solar magnetic field and drive the FBI with
enough energy to account for upper chromospheric heating. This would
explain the radiative losses in both the quiet-Sun internetwork and
network lanes. These papers applied an oversimplified approach to
the FBI appropriate for the ionosphere but needs some modification
to accurately model the instability in the metal-ion dominated chromosphere.
The current paper derives a chromospheric FBI which modifies the results
of Fontenla (2005) and Fontenla et al. (2008), lending further support
for their conclusions.

The FBI was first used to explain density irregularities in the equatorial
electrojet observed in ionospheric radar experiments (Farley 1963;
Buneman 1963). The instability occurs in weakly ionized, collisional
plasmas with strongly magnetized electrons but collisionally demagnetized
ions. Electrostatic waves develop due to the disparate motions of
the electrons and ions when plasma flows across magnetic field lines.
The zeroth order motion of the electrons is the $\boldsymbol{E}\times\boldsymbol{B}$
(Hall) drift while the ions necessarily follow the neutral flow because
of the high ion-neutral collision rates. Linear wave growth develops
if the electron $\boldsymbol{E}\times\boldsymbol{B}$ drift velocity
exceeds the ion-acoustic velocity multiplied by a dimensionless factor
close to unity. The instability in the ionosphere develops on a time
scale somewhat larger than the ion-neutral mean free time, which means
it operates on a millisecond to 10s of millisecond time scale. This
exceeds Alfvénic time scales but is below most electron frequencies.

The electric field that creates the $\boldsymbol{E}\times\boldsymbol{B}$
drift that drives the FBI in the ionosphere arises from different
causes in various regions of the Earth's E-region ionosphere. Near
the magnetic equator, the predominant energy source for this derives
from strong neutral winds that flow East-West across the largely horizontal
North-South pointing geomagnetic field (Richmond 1973). Since the
ions must follow the neutrals but the electrons remain mostly tied
to field lines, a complex current develops, and that, combined with
the vertical gradient in the neutral density and hence the conductivity,
causes the formation of strong East-West currents called electrojets.
In the Earth's auroral regions, where radars detect ferociously strong
FBI waves, electric fields generated in the coupled magnetosphere-ionosphere
system propagate down the mostly vertical magnetic fields (Dimant
\& Oppenheim 2010a; 2010b). These fields drive auroral electrojets
with hypersonic electron flow rates and the FBI then heats the electrons
in this region dramatically (Foster \& Erickson 2000; Oppenheim \&
Dimant 2013).

This paper expands the theory of the FBI employed by Fontenla (2005),
Fontenla et al. (2008), and Gogoberidze et al. (2009) to account for
the diversity of ion species in the solar chromosphere. The single
species theory is appropriate for the E-region ionosphere where the
primary ion constituents are O$_{2}^{+}$ and NO$^{+}$ which are
similar in mass and can be treated as a single species. However, ionic
components of chromospheric plasma range from protons to ionized metals
dominated by Si II, Mg II, and Fe II. Assuming an average ion mass
ignores the physics arising from differences in mobilities and collision
frequencies among the various ion species. This paper derives the
linear, multi-species dispersion relation and electron $\boldsymbol{E}\times\boldsymbol{B}$
drift trigger velocity appropriate for chromospheric conditions. It
then applies the multi-species theory to a recent model of the solar
chromosphere to determine the likely range of electron drift velocities
necessary to trigger the instability as a function of pressure within
the chromosphere.

\section{Theory}

\subsection{Linear Dispersion Relationship and Trigger Velocity}

A model of the FBI applicable to the chromospheric plasma requires
deriving a multi-species dispersion relationship using a linear, fluid
approximation similar to that used by Farley (1963) and Buneman (1963),
and later refined by Sudan et al. (1973). To obtain the simplest manageable
dispersion relationship, we will make several assumptions. First,
we assume all ion species are demagnetized and electrons are magnetized.
A species is magnetized when its gyrofrequency greatly exceeds its
collision frequency. Given the temporal and spatial scales under consideration,
we approximate the electric field with an electrostatic field represented
by the gradient of a scalar potential. Next, we take all ions to be
singly ionized and assume the plasma is quasineutral since we are
concerned with frequencies much smaller than the plasma frequency.
We consider electron inertia negligible due the large difference in
mass between electrons and any ion species. Also, we consider the
ion Pedersen drift negligible allowing the ions to be at rest in the
frame of the neutral flow to zeroth order. Furthermore, we take the
zeroth order electron velocity to be its $\boldsymbol{E}\times\boldsymbol{B}$
drift velocity.

Assuming linear, plane wave perturbations, wave propagation will occur
predominantly in the direction perpendicular to $\boldsymbol{B}$;
in other words, $k_{\shortparallel}^{2}\ll k_{\bot}^{2}$ where $\boldsymbol{k}_{\shortparallel}$
and $\boldsymbol{k}_{\perp}$ are the wavevector components parallel
and perpendicular to $\boldsymbol{B}$, respectively. Only long wavelength
waves are considered; in other words, $kv_{D}\ll\nu_{j}$ where $v_{D}$
is the electron $\boldsymbol{E}\times\boldsymbol{B}$ drift velocity
and $\nu_{j}$ is the total ion momentum transfer collision frequency
for the j$^{\text{th}}$ ion species (Dimant \& Oppenheim 2004). However,
these wavelengths are much longer than those characteristic of Alfvénic
modes. We assume the plasma has no zeroth order density gradients
which could greatly enhance the instability but will lead to a far
more complex analysis.

We start our analysis with the fluid equations of motion for both
electrons and ions:

\begin{equation}
n\equiv n_{e}=\sum_{j}n_{j}\label{eq:quasiN}
\end{equation}
\begin{equation}
\frac{\partial n}{\partial t}+\boldsymbol{\nabla}\cdot(n\boldsymbol{v_{e}})=0\label{eq:eCont}
\end{equation}
\begin{equation}
\frac{\partial n_{j}}{\partial t}+\boldsymbol{\nabla}\cdot(n_{j}\boldsymbol{v}_{j})=0
\end{equation}
\begin{equation}
e\left(\boldsymbol{\nabla}\phi-\boldsymbol{v}_{e}\times\boldsymbol{B}\right)-\frac{\gamma_{e}K_{B}T_{e}\mathbf{\boldsymbol{\nabla}}n}{n}-m_{e}\nu_{e}\boldsymbol{v}_{e}=0\label{eq:elecMoment}
\end{equation}
\begin{equation}
m_{j}\left[\frac{\partial\boldsymbol{v}_{j}}{\partial t}+\left(\boldsymbol{v_{j}}\cdot\boldsymbol{\nabla}\right)\boldsymbol{v_{j}}\right]=-e\boldsymbol{\nabla}\phi-\frac{\gamma_{j}K_{B}T_{j}\boldsymbol{\nabla}n_{j}}{n_{j}}-m_{j}\nu_{j}\boldsymbol{v_{j}}\label{eq:ionMoment}
\end{equation}
where $n$ is number density, $\boldsymbol{v}$ is flow velocity,
$e$ is the elementary charge, $\phi$ is electrostatic potential,
$\boldsymbol{B}$ is the magnetic field, $\gamma$ is the ratio of
heat capacities, $K_{B}$ is the Boltzmann constant, $T$ is temperature,
$m$ is mass, and the subscript $e$ corresponds to electrons while
the subscript $j$ corresponds to the j$^{\text{th}}$ ion species.
We maintain separate temperatures for each species even though collisions
will keep the ion temperatures quite similar to the neutral ones.
The electron temperature, however, can become somewhat elevated by
a range of processes. The collision rates, $\nu_{e}$ and $\nu_{j}$,
include both collisions with neutrals and coulomb collisions. A full
treatment of Coulomb collisions would add a few additional components
to eqs (\ref{eq:elecMoment}) and (\ref{eq:ionMoment}), but these
only modify our final results by a negligible amount (Gogoberidze
et al. 2009). This restricts our analysis to near the temperature
minimum where collisions with neutrals dominate over Coulomb collisions. 

To find the dispersion relationship, we keep only the first-order,
linear terms in eqs (\ref{eq:quasiN})-(\ref{eq:ionMoment}). We assume
plane wave perturbations such that all dynamical quantities vary as
$\xi=\delta\xi\exp[-i(\omega t-\boldsymbol{k}\cdot\boldsymbol{x})]$,
providing:
\begin{equation}
\left(\omega-\boldsymbol{k}\cdot\boldsymbol{v}_{D}\right)\delta n-n_{0}\left(\boldsymbol{k}\cdot\boldsymbol{\delta v_{e}}\right)=0
\end{equation}
\begin{equation}
\omega\delta n_{j}-n_{j0}\left(\boldsymbol{k}\cdot\boldsymbol{\delta v_{j}}\right)=0
\end{equation}
\begin{equation}
e\left(i\boldsymbol{k}\delta\phi-\boldsymbol{\delta v_{e}}\times\boldsymbol{B}\right)-i\frac{\gamma_{e}K_{B}T_{e}}{n_{0}}\boldsymbol{k}\delta n-m_{e}\nu_{e}\boldsymbol{\delta v_{e}}=0
\end{equation}
\begin{equation}
m_{j}\left(\omega+i\nu_{i}\right)\boldsymbol{\delta v_{j}}-e\boldsymbol{k}\delta\phi-\frac{\gamma_{j}K_{B}T_{j}}{n_{j0}}\boldsymbol{k}\delta n_{j}=0
\end{equation}
\begin{equation}
\delta n=\sum_{j}\delta n_{j}
\end{equation}
where the subscript 0 represents zeroth order quantities, and $\delta$
represents a linearly perturbed Fourier coefficient. This analysis
is in the neutral fluid rest frame where the electrons have a zero-order
velocity of $\boldsymbol{v_{D}}\equiv\boldsymbol{E}\times\boldsymbol{B}/|\boldsymbol{B}|$
and the small ion Pedersen drifts are inconsequential. 

Eliminating all of the linearly perturbed Fourier coefficients yields
the dispersion relationship:
\begin{equation}
D(\omega,\boldsymbol{k})\equiv\left[\omega-\boldsymbol{k}\cdot\boldsymbol{v_{D}}+i\frac{\nu_{e}k^{2}U_{e}^{2}}{\Omega_{e}^{2}}\left(1+\frac{\Omega_{e}^{2}k_{\shortparallel}^{2}}{\nu_{e}^{2}k^{2}}\right)\right]\sum_{j}\frac{n_{j0}}{n_{0}\psi_{j}}\left(\omega-i\frac{\omega^{2}}{\nu_{j}}+i\frac{k^{2}U_{j}^{2}}{\nu_{j}}\right)^{-1}+1=0\label{eq:multSpDisp}
\end{equation}
where $\Omega$ is the gyrofrequency, $U_{j}\equiv\sqrt{\gamma_{j}K_{B}T_{j}/m_{j}}$
is the thermal velocity, and $\psi_{j}$ is defined as:
\begin{equation}
\psi_{j}\equiv\frac{\nu_{j}\nu_{e}}{\Omega_{j}\Omega_{e}}\left(1+\frac{\Omega_{e}^{2}k_{\shortparallel}^{2}}{\nu_{e}^{2}k^{2}}\right).
\end{equation}
Assuming only one ion species reduces eq (\ref{eq:multSpDisp}) to
the single-ion dispersion relation found in Sudan et al. (1973):
\begin{equation}
\omega\left(1+\psi_{j}\right)-\boldsymbol{k}\cdot\boldsymbol{v}_{D}=i\frac{\psi_{j}}{\nu_{j}}\left[\omega^{2}-\left(\frac{m_{e}}{m_{j}}U_{e}^{2}+U_{j}^{2}\right)k^{2}\right].
\end{equation}

Returning to eq (\ref{eq:multSpDisp}), let $\omega\equiv\omega_{r}+i\Gamma$,
where $\omega_{r}$ represents the oscillation frequency of the waves
and $\Gamma$ represents the growth/damping rate of the waves. To
recover a phase-velocity relation from (\ref{eq:multSpDisp}), we
will ignore all small and imaginary terms by taking $\Gamma\rightarrow0$
and $\omega_{r}/\nu_{j}\rightarrow0$ for all ion species $j$. This
provides:
\begin{equation}
\omega_{r}=\frac{\boldsymbol{k}\cdot\boldsymbol{v}_{D}}{1+\psi}\label{eq:oscFreq}
\end{equation}
where
\begin{equation}
\psi\equiv\left(\sum_{j}\frac{n_{j0}}{n_{0}\psi_{j}}\right)^{-1}
\end{equation}
which is equivalent to the oscillation frequency from Sudan et al.
(1973). Now, we will find an approximate solution for the growth rate
by assuming slow growth ($|\Gamma|\ll|\omega_{r}|$) and expanding
(\ref{eq:multSpDisp}) about $\Gamma=0$:
\[
D(\omega,\boldsymbol{k})=Re\left(D\left(\omega,\boldsymbol{k}\right)\right)+iIm\left(D\left(\omega,\boldsymbol{k}\right)\right)=0
\]
\[
Re(D(\omega_{r},\boldsymbol{k}))+i\Gamma\frac{\partial Re(D(\omega,\boldsymbol{k}))}{\partial\omega}|_{\omega=\omega_{r}}\approx-iIm(D(\omega_{r},\boldsymbol{k}))
\]
\begin{equation}
\Gamma\approx-Im(D(\omega_{r},\boldsymbol{k}))/\frac{\partial Re(D(\omega,\boldsymbol{k}))}{\partial\omega}|_{\omega=\omega_{r}}.\label{eq:growthGen}
\end{equation}
Separating eq (\ref{eq:multSpDisp}) into real and imaginary parts
and substituting into eq (\ref{eq:growthGen}) provides the growth/damping
rate:
\begin{equation}
\Gamma\approx\frac{\psi^{2}}{1+\psi}\sum_{j}\frac{n_{j0}}{n_{0}\psi_{j}\nu_{j}}\left[\omega_{r}^{2}-\left(\frac{m_{e}}{m_{j}}U_{e}^{2}+\frac{\psi_{j}}{\psi}U_{j}^{2}\right)k^{2}\right].\label{eq:growthApp}
\end{equation}
The electron drift trigger velocity above which the instability will
occur can be found by setting eq (\ref{eq:growthApp}) to zero, assuming
an optimal direction, $\boldsymbol{k}\parallel\boldsymbol{v_{D}}$,
and substituting for the oscillation frequency using (\ref{eq:oscFreq}):
\begin{equation}
v_{trig}=\left(1+\psi\right)\sqrt{\sum_{j}\frac{n_{j0}}{\psi_{j}\nu_{j}}\left(\frac{m_{e}\psi_{j}}{m_{j}\psi}U_{e}^{2}+U_{j}^{2}\right)/\sum_{j}\frac{n_{j0}}{\psi_{j}\nu_{j}}}.\label{eq:multSpec}
\end{equation}
This expression significantly differs from the expression for the
'single species' threshold velocity, $v_{trig}=\left(1+\psi\right)c_{I}$,
where $c_{I}$ is the ion-acoustic velocity defined as:
\begin{equation}
c_{I}^{2}\equiv\sum_{j}\frac{n_{j0}}{n_{0}}\left(\frac{m_{e}}{m_{j}}U_{e}^{2}+U_{j}^{2}\right).
\end{equation}
In particular, eq (\ref{eq:multSpec}) replaces the ion-acoustic speed
in the single ion trigger velocity equation with a term that couples
both thermal and collisional phenomena. From this, we expect the multi-species
trigger velocity to be more sensitive to ion-neutral, electron-neutral,
and electron-ion collision frequencies than the single species trigger
velocity where all of the collisional information is contained within
the typically small $\psi$ coefficient.

\subsection{Collision Frequencies}

The trigger velocity of the FBI depends sensitively on the collision
rates. In this analysis, we consider three types of elastic collisions:
momentum transfer collisions between charged species and neutral hydrogen,
resonant charge exchange between protons and neutral hydrogen, and
Coulomb collisions between ions and electrons.

First, we consider momentum transfer collisions experienced by all
charged species with neutrals. Unfortunately, little experimental
data about elastic collisions between neutral hydrogen and metal ion
species for energies between 0.1 eV and 1 eV exist. Instead, we must
rely upon a relatively simple collision model that can be generalized
to any ion species in the solar chromosphere. To do this, we assume
the dominant scattering mechanism is the repulsive force provided
by the dipole polarization of neutral hydrogen when subjected to the
electric field produced by the ion species. For charged species \emph{s}
in the neutral frame, Dalgarno et al. (1958) provides this classical
result for the momentum transfer collision frequency with a neutral
species \emph{n}:
\begin{equation}
\nu_{sn}=2.21\pi n_{n}\sqrt{\frac{\alpha_{n}e^{2}}{4\pi\epsilon_{0}\mu_{sn}}}\label{eq:CollisionFreq}
\end{equation}
where $n_{n}$ is the number density of the neutral species, $\alpha_{n}$
is the polarizability of the neutral species, $\epsilon_{0}$ is the
vacuum permittivity, and $\mu_{sn}$ is the reduced mass of the charged
and neutral species. By neglecting quantum effects, this relationship
underestimates the collision rates for metal ions.

We can gauge how much eq (\ref{eq:CollisionFreq}) underestimates
the collision frequencies by looking at the relatively few sophisticated
calculations performed for some of the lighter metal species at 0.1
eV to 1 eV. Kristi\'{c} and Schultz (2009) and Liu et al. (2010) calculated
momentum transfer collision frequencies between Be II and C II with
neutral hydrogen using semi-classical and quantum methods. Although
Be II is extremely rare in the solar chromosphere, we can still use
it to determine the validity of eq (\ref{eq:CollisionFreq}). Within
the energy range of 0.1 eV to 1 eV for Be II and C II, eq (\ref{eq:CollisionFreq})
underestimates the results of Kristi\'{c} \& Schultz (2009) and Liu
et al. (2010) by 70\% at the greatest. These calculations are more
difficult to perform for massive ion species with complicated electronic
structures. However, we expect increases in the polarization collision
frequencies of heavy ions (e.g. Si II, Mg II, Fe II) to have a minimal
effect upon eq (\ref{eq:multSpec}), the trigger velocity. For example,
increasing the collision frequencies for heavy ions by a factor of
two only increases our results for eq (\ref{eq:multSpec}) by 20\%
which is not significant enough to factor into the conclusions of
this paper. In light of this, we will apply equation (\ref{eq:CollisionFreq})
to all charged species.

Resonant charge exchange occurs when an ion receives an electron from
the neutral with a similar ionization energy, effectively switching
the roles of the two. This converts fast ions into fast neutrals and
vice versa efficiently enough to still be considered elastic collisions
for the purpose of this analysis. Resonant charge exchange becomes
a significant collision mechanism between protons and neutral hydrogen
at temperatures above 300 K (Banks \& Kockarts 1973). Using a Maxwellian
averaged cross-section, Schunk \& Nagy (2009) provides:
\begin{equation}
\nu_{res}=2.65\times10^{-16}n_{H}\sqrt{\frac{T_{H}+T_{protons}}{2}}\left[1-0.083\log_{10}\left(\frac{T_{H}+T_{protons}}{2}\right)\right]^{2}\label{eq:ResCollisionFreq}
\end{equation}
where $n_{H}$ has units of m\textsuperscript{-3}.

Coulomb collisions between ions and electrons become important at
higher altitudes in the solar chromosphere and can increase the likelihood
of the FBI forming there (Gogoberidze et al. 2009). The Coulomb collision
frequency for charged species \emph{s} in the neutral frame, including
the Spitzer correction for small angle collisions, is:
\begin{equation}
\nu_{Coul,s}=\frac{\pi n_{e}e^{4}\ln\left(12\pi n_{e}\lambda_{D}\right)}{\left(4\pi\epsilon_{0}\right)^{2}\sqrt{m_{s}\left(2K_{B}T\right)^{3}}}\label{eq:CoulCollisionFreq}
\end{equation}
where $\lambda_{D}\approx\sqrt{\epsilon_{0}K_{B}T/n_{e}e^{2}}$ is
the Debye length. 

The total collision frequency for each charged species, $\nu_{s}$,
is the direct sum of eqs (\ref{eq:CollisionFreq}) and (\ref{eq:CoulCollisionFreq}).
For protons, we add eq (\ref{eq:ResCollisionFreq}) to the sum. Figure
2 shows the collision frequencies normalized by the gyrofrequency
as a function of pressure within the solar atmosphere.

\section{Results and Discussion}

Combining the multi-species trigger velocity predicted by eq (\ref{eq:multSpec})
and a semi-empirical, one-dimensional, NLTE model of the solar chromosphere
developed by Fontenla et al. (2009) allows one to calculate the trigger
velocity for the FBI as a function of pressure. The model provides
densities and temperatures for all abundant ion species from hydrogen
to zinc as functions of pressure. The dominant ion species just below
the temperature minimum are Mg II, Si II, and Fe II while protons,
and to a lesser extent C II, dominate above as shown in Figure 3.
The plasma just below the temperature minimum is unlike any observed
in the geospace environment because of the high masses and low mobilities
of its dominant ion species. As one moves up from the photosphere,
the plasma cools, and the protons and electrons in the plasma begin
to recombine leaving a plasma dominated by metals. Before one gets
to the temperature minimum, the trend reverses as a result of ionizing
radiation from above as demonstrated in Figure 3. Though the Fontenla
et al. (2009) model makes detailed predictions, the exact altitude
where these changes occur in reality remains uncertain since the model
is one-dimensional and has a number of uncertainties.

Figure 4 shows the multi-species trigger velocity as a function of
pressure over a range of magnetic field strengths. For this analysis,
we adopt a magnetic field range of 30 G to 120 G. This range follows
from investigations of the Hanle effect in atomic and molecular lines
(Trujillo Bueno et al. 2004; 2006). Near the base of the chromosphere,
the large neutral density demagnetizes the electrons, producing a
large $\psi$ coefficient and preventing the instability from forming.
As one moves upward, the trigger velocity rapidly approaches a minimum
value near the temperature minimum. The velocity minimum migrates
lower into the chromosphere when magnetic field strength increases
due to stronger electron magnetization. The multi-species trigger
velocity achieves a minimum value of about 7 km s\textsuperscript{-1}
for 30 G magnetic fields and 4 km s$^{\text{-1}}$ for 120 G fields.
Note that for the largest estimates of magnetic field strength ($\gtrsim75$
G), the model predicts trigger velocities below the neutral acoustic
speed. This is significant since one never finds subsonic trigger
velocities in the ionosphere, and despite these higher thresholds,
the ionospheric FBI is ubiquitous there. Convective overshoots from
granular and supergranular flows in the photosphere will drag plasma
across magnetic field lines near the temperature minimum. The kinetic
energy of these flows should provide enough free energy to induce
strong electric fields, producing electron drift velocities sufficient
to trigger the FBI. This may also provide enough energy to maintain
the persistent and global heating contributed by the instability.
As one continues upward above the temperature minimum, the trigger
velocity increases to approximately 10 km s$^{\text{-1}}$ for all
magnetic field strengths between 30 G and 120 G. This increase in
trigger velocity makes the instability more difficult to drive, though
given the copious amounts of free energy available, it may continue
to play a role at high altitudes. A dramatic increase in proton density
is responsible for the higher thresholds at these altitudes.

Eq (\ref{eq:multSpec}) assumes strong electron magnetization and
ion demagnetization. Given the substantial range of magnetic fields
that exist in the chromosphere, we can check the validity of these
assumptions. Figure 2 plots the collision frequency to gyrofrequency
ratios for electrons, protons, and Fe II which we use to represent
the behavior of all metallic ion species. Near the temperature minimum,
electrons appear strongly magnetized while Fe II is strongly demagnetized
for all magnetic field strengths. However, the proton frequency ratio
ranges from 0.1 to 10, indicating that protons may potentially magnetize
near the temperature minimum. If protons magnetize near the temperature
minimum where heavy ions dominate, it may not eliminate the instability,
but rather enhance it. Magnetization could limit proton mobility across
magnetic field lines, promoting wave growth since the protons would
have a more difficult time shorting out the instability in regions
where heavy, metallic ions drive the instability. As one moves further
up, protons overwhelmingly dominate and become magnetized, shutting
off the instability; however, the our predictions from eq (\ref{eq:multSpec})
do not yet include this physics.

\section{Conclusions}

The multiple ion FBI theory developed above predicts an electron $\boldsymbol{E}\times\boldsymbol{B}$
drift trigger velocity below the neutral acoustic speed for magnetic
field strengths greater than 75 G. In the Earth's auroral electrojet
this instability causes intense electron heating and should do the
same in the chromosphere. In the chromosphere, the FBI would develop
near the solar temperature minimum and extinguish in the upper chromosphere
due to the shift from a metal-dominated to a proton-dominated ion
population.

Both simulations (Nordlund et al. 1997; Vögler et al. 2005; Freytag
et al. 2012) and Hinode observations (Rieutord et al. 2010) show that
granule motions at the top of the photosphere drive intense ``neutral
winds'' across solar magnetic field lines. In the Earth's equatorial
ionosphere, cross-field neutral winds generate electric fields that
easily exceed the Earth's FBI threshold on a daily basis. One would
expect that the strong solar flows would create intense electric fields
which should easily drive the FBI. As in the Earth's ionosphere, electric
fields driven by neutral flows in the solar atmosphere will travel
along magnetic field lines; hence, a strong wind across $\boldsymbol{B}$
below the region where the FBI has the lowest threshold may propagate
to this region and trigger the instability. Likewise, as in the auroral
ionosphere, energy entering the chromosphere from above or below,
possibly in the form of Alfvén waves, will manifest itself as strong
drifts and electric fields that can trigger the FBI. In the Earth's
electrojet, the trigger velocities exceed the both the neutral acoustic
and the highest wind speeds but the FBI still occurs because the driving
electric field exceeds the local $\boldsymbol{E}=-\boldsymbol{u}\times\boldsymbol{B}$
fields due to a variety of mechanisms. Similarly amplified electric
fields may not be necessary to trigger the FBI in the chromosphere,
though they may exist.

Directly observing these electric fields will prove difficult, though
indirect observations may be possible. In the ionosphere, few in situ
measurements of electric fields exist; however, numerous indirect
measurements exist (Foster \& Erickson 2000). Near the Earth's electrojets,
the field is inferred from ground-based magnetometer measurements
since the electric field drives Hall currents which cause easily detected
perturbations of the geomagnetic field (such changes were among the
earliest space-physics measurements, dating to the 1700s). More precise
measurements of electric fields are made by radars in the auroral
electrojet by detecting field-induced $\boldsymbol{E}\times\boldsymbol{B}$
drifts on field lines above, but connected to, the E-region. At these
higher altitudes, electrons and ions drift together due to reduced
collision rates. A combination of detailed observations and modeling
should allow us to estimate the magnitude of the plasma drifts and
fields in the chromosphere. 

In this work, we ignored the effects of proton magnetization on the
multi-species FBI in the solar chromosphere. Unlike heavier and less
mobile ions, protons principally act to short out the perturbed electric
fields driving the instability. Magnetizing protons would limit their
mobility across magnetic field lines, potentially enhancing the instability.
We plan to analyze this more fully in a future work.

This paper further develops the ideas first put forward in Fontenla
(2005) and Fontenla et al. (2008) by more accurately modeling the
triggering of the FBI in the chromosphere. The argument that this
instability should exist in the chromosphere and that it has the potential
of being an important mechanism for electron heating in the these
regions remains robust. It will operate on rapid time scales much
faster than the minute time scales typically considered by MHD systems.
Also, its driver will be the neutral flows in the upper photosphere
and lower chromosphere, making it a good candidate for heating of
the quiet Sun.

\acknowledgements{This work was supported in part by NSF through Ionospheric Physics
Grants Nos. ATM-0442075, ATM-0819914, and ATM-1007789. The authors
would like to thank Prof. Javier Trujillo Bueno for bring an error
to their attention.}

\pagebreak{}

\section{Figures}

\noindent \includegraphics[clip]{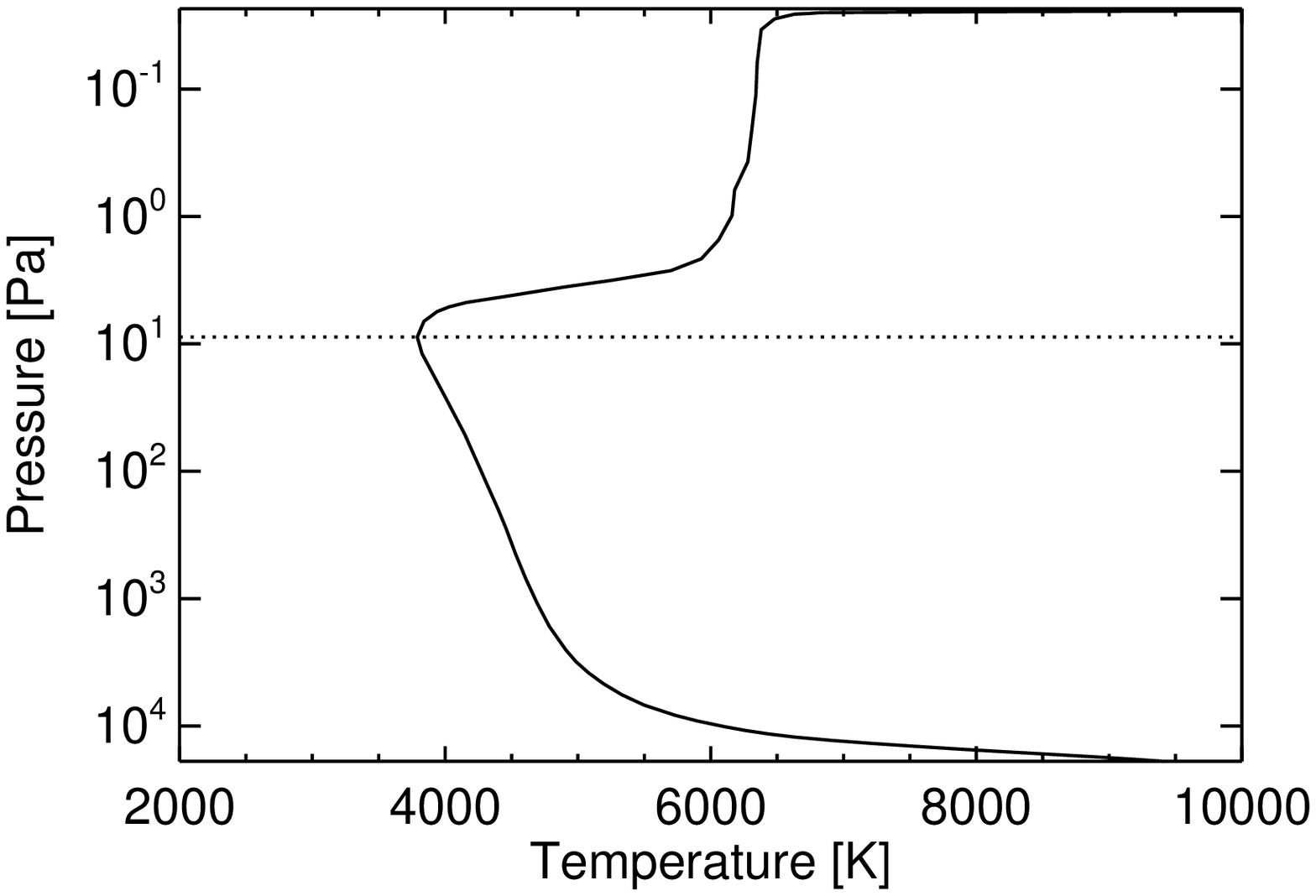}

\figcaption{Temperature profile of the solar chromosphere from the Fontenla (2009)
model. The solar temperature minimum is marked by the dotted line.
The temperature increases sharply by over 2000 K immediately above
the temperature minimum.}

\pagebreak{}

\noindent \includegraphics[clip]{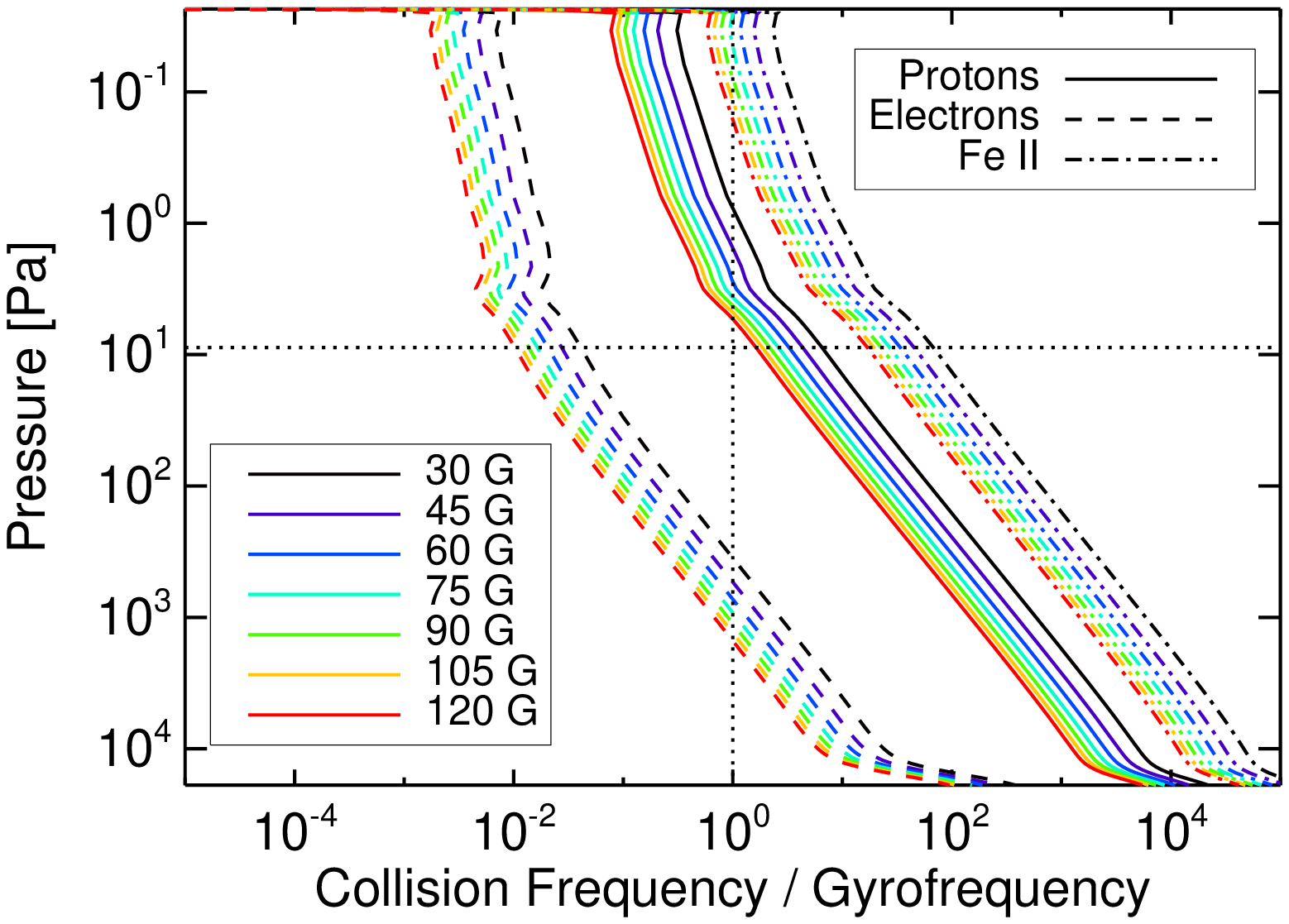}

\figcaption{Collision frequency normalized to the gyrofrequency for protons,
electrons, and Fe II over a range of magnetic field strengths from
30 G to 120 G. Values greater than unity imply a demagnetized species.
Near the temperature minimum, electrons are strongly magnetized while
heavy ions (represented by Fe II) are strongly demagnetized. However,
protons may become magnetized near the temperature minimum, potentially
enhancing the instability there.}

\pagebreak{}

\noindent \includegraphics[clip]{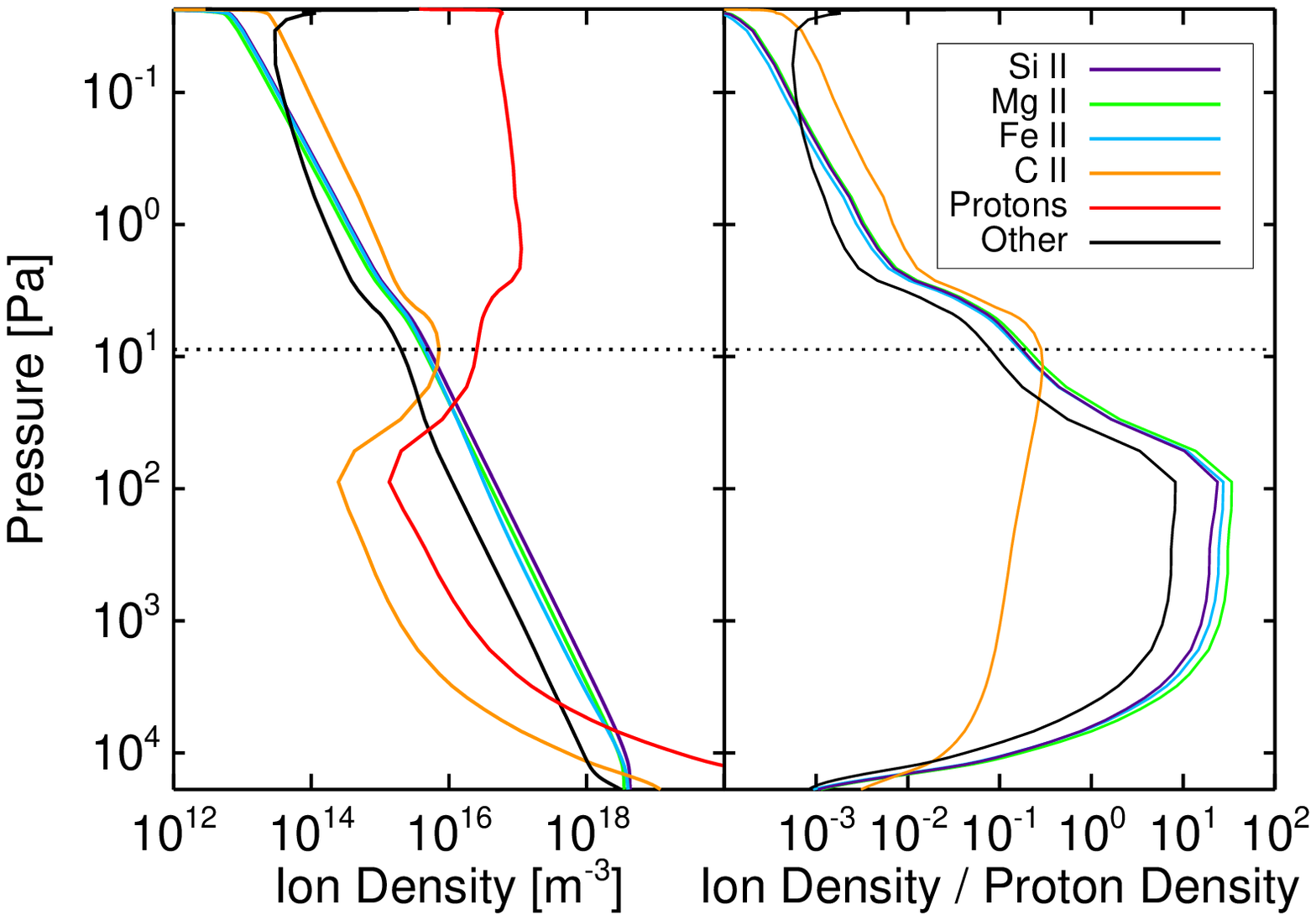}

\figcaption{Left plot: ion density profiles for Si II (purple), Mg II (green),
Fe II (blue), C II (orange), protons (red), and remaining ion species
(black) from Fontenla et al. (2009). Right plot: ratios of ion density
to proton density for the same ion species. Massive metallic ions
(Si II, Mg II, Fe II) dominate below the temperature minimum, while
protons, and to a lesser extent C II, dominate above the temperature
minimum.}

\pagebreak{}

\noindent \includegraphics[clip]{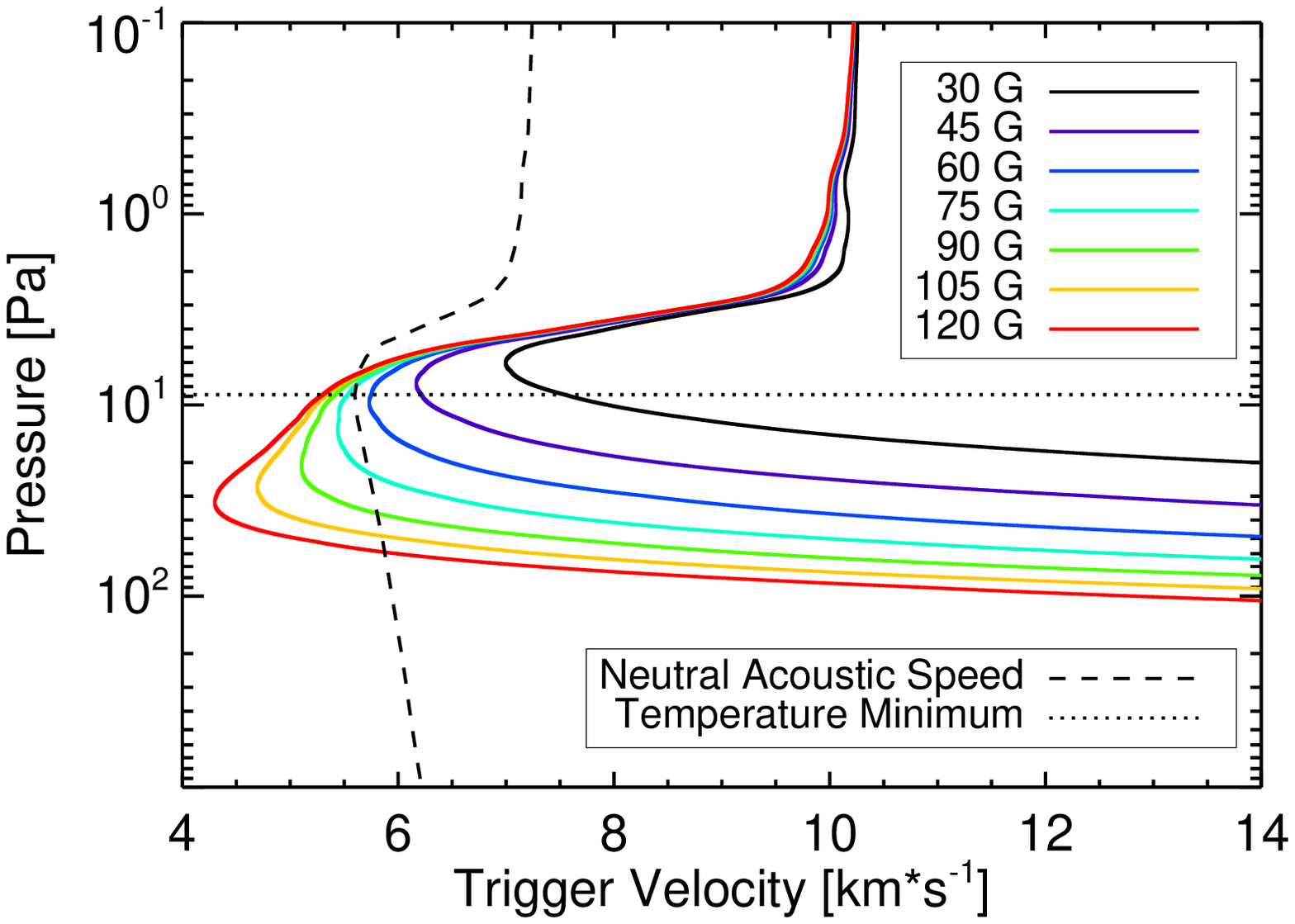}

\figcaption{Predicted multi-species trigger velocity at several magnetic field
strengths. The plot is limited to pressures near the temperature minimum
where eq (\ref{eq:multSpec}) is valid. The trigger velocity decreases
sharply as it approaches the temperature minimum reaching speeds as
low as 4 km s\textsuperscript{-1} at the largest magnetic field strengths.
The neutral acoustic speed is plotted for reference.}

\pagebreak{}

\end{document}